\documentclass{emulateapj} 
 
\usepackage[dvipdf]{epsfig} 
\usepackage[dvips]{rotating}
\usepackage{subfigure}
\usepackage{graphicx}

\newcommand{\IRAS}{{\it IRAS}}
\newcommand{\HERSCHEL}{{\it Herschel}}
\newcommand{\SPITZER}{{\it Spitzer}}

\bibpunct{(}{)}{;}{a}{}{,} 
 
\shorttitle{WISE 12 Micron Atlas}
 
\shortauthors{Meisner \& Finkbeiner} 
 
\begin{document} 
\title{A Full-sky, High-resolution Atlas of Galactic 12 Micron Dust Emission 
with WISE}

\author{Aaron M. Meisner\altaffilmark{1,2}}
\author{Douglas P. Finkbeiner\altaffilmark{1,2}}
\altaffiltext{1}{Department of Physics, Harvard University, 17 Oxford Street, 
Cambridge, MA 02138, USA; ameisner@fas.harvard.edu}
\altaffiltext{2}{Harvard-Smithsonian Center for Astrophysics, 60 Garden St, 
Cambridge, MA 02138, USA; dfinkbeiner@cfa.harvard.edu}

\begin{abstract} 
We describe our custom processing of the entire Wide-field Infrared Survey 
Explorer (WISE) 12$\mu$m imaging data set, and present a 
high-resolution, full-sky map of diffuse Galactic dust emission that is 
free of compact sources and other contaminating artifacts. The principal 
distinctions between our resulting co-added images and the WISE Atlas stacks 
are our removal of compact sources, including their associated electronic and 
optical artifacts, and our preservation of spatial modes larger than 
$1.5^{\circ}$. We provide access to the resulting full-sky map via a set of 
430 12.5$^{\circ} \times$12.5$^{\circ}$ mosaics. These stacks have been 
smoothed to 15$''$ resolution and are accompanied by corresponding coverage 
maps, artifact images, and bit-masks for point sources, resolved compact 
sources, and other defects. When combined appropriately with other 
mid-infrared and far-infrared data sets, we expect our WISE 12$\mu$m co-adds 
to form the basis for a full-sky dust extinction map with angular resolution 
several times better than \cite{SFD}.
\end{abstract}  
 
 
\section{Introduction}

High-latitude complexes of resolved infrared sources 
were first identified by \cite{low84} using \IRAS~60$\mu$m and 
100$\mu$m observations \citep{wheelock94}. This so-called ``infrared cirrus,'' 
attributable 
to thermal emission from large dust grains in the interstellar medium, has 
since been probed in greater detail by a variety of instruments and 
detected over virtually the entire sky. Beginning in 1989, DIRBE 
 mapped the full sky at ten infrared wavelengths from 1.25$\mu$m to 240$\mu$m 
with a reliable zero point, but inferior 0.7$^{\circ}$ angular resolution 
\citep{boggess92}. Later, the ISO and IRTS missions 
\citep{iso96, onaka96, tanaka96} established that 
polycyclic aromatic hydrocarbons (PAHs) are responsible for the 
presence of cirrus at mid-IR wavelengths, displaying strong emission bands at 
8.6$\mu$m, 11.3$\mu$m, and 12.6$\mu$m. Present day observatories are now 
capable of mapping the mid-IR and far-IR cirrus with excellent resolution and 
sensitivity. For example, \HERSCHEL~has conducted pointed imaging observations 
from 70$\mu$m to 500$\mu$m with corresponding resolution ranging from 5.2$''$ 
to 37$''$ \citep{herschel}. However, \HERSCHEL~and \SPITZER~\citep{spitzer}, 
each with field-of view $\lesssim$5$'$, only observed a very small fraction
of the sky over their lifetimes.

On the other hand, the Wide-field Infrared Survey Explorer 
\citep[WISE,][]{wright10} has imaged the full sky in four broad infrared 
bandpasses at high angular resolution \citep{cutri12}. In the two 
bluest channels ($W1 = 3.4\mu$m, $W2 = 4.6\mu$m), the Rayleigh-Jeans tail of 
stellar emission dominates, with 
images displaying vast numbers of compact sources, and diffuse dust structure 
that is typically discernible only at very low Galactic latitudes. At 
the other end, the reddest band ($W4 = 22\mu$m) is strongly dominated by 
our Solar System's interplanetary dust, its thermal emission peaking between 
15$\mu$m and $25\mu$m. While the zodiacal foreground is still bright at 
$W3 = 12\mu$m, this bandpass was chosen to coincide with the strongest 
PAH emission features. As a result, Galactic cirrus is often 
prominent in $W3$, and has been imaged by WISE with high signal-to-noise over 
most of the sky in spite of bright zodiacal contamination that varies on large 
angular scales. Further, with a native resolution of $6.5''$, WISE $W3$ 
represents a dramatic improvement in angular resolution relative to previous 
mid-IR and far-IR full-sky data sets. This has 
motivated us to investigate WISE 12$\mu$m images as a way of mapping the 
diffuse mid-IR emission from Galactic interstellar dust and probing small-scale
structure in the Milky Way ISM. 


However, official WISE mission imaging data products do not include
a clean map of the 12$\mu$m Galactic cirrus. To isolate
diffuse $W3$ nebulosity, we have undertaken a custom per-exposure 
processing of the $W3$ data set and constructed custom stacks from these 
processed frames. In particular, we have:
 
$\bullet$ subtracted point sources and artifacts of bright point \indent \ \ sources

$\bullet$ masked moving Solar System objects

$\bullet$ masked resolved compact sources


$\bullet$ eliminated single-epoch anomalies (e.g. satellite

 \ \ streaks and cosmic rays)

$\bullet$ corrected single-frame striping artifacts and per-

 \ \ quadrant offsets

$\bullet$ addressed time-dependent foregrounds due to the 

 \ \ Moon and zodiacal light

$\bullet$ preserved large angular scale modes

Because PAHs trace the far-IR thermal emission from 
large grains, a particularly interesting application of diffuse $W3$ maps is 
the creation of high-resolution foreground templates and extinction estimates. 
In terms of predicting dust extinction, the industry standard is 
\citet[hereafter SFD]{SFD}, with 6.1$'$ angular resolution limited by the 
asymmetric 4$'$ \IRAS~beam. Even if we must smooth $W3$ 
considerably relative to its native resolution, we can still characterize the 
distribution of interstellar dust over the full sky on angular scales many 
times smaller than the SFD/\IRAS~beam. 

In $\S$\ref{sec:data} we summarize the relevant details of the WISE survey 
and its data products. In $\S$\ref{sec:ptsrc}, we describe 
the point source catalog which enables us to model the PSF and perform PSF 
subtractions. In $\S$\ref{sec:psf} we discuss our model of the PSF and related 
bright source artifacts. In $\S$\ref{sec:l1b} we describe our single-exposure 
processing pipeline that eliminates many contaminants of the diffuse $W3$ 
emission. In $\S$\ref{sec:stack} we discuss our custom stacks built 
from these processed single exposures. In $\S$\ref{sec:tiles} we discuss the 
full-sky map we are releasing\footnote{See http://wise.skymaps.info.} and conclude in $\S$\ref{sec:disc}.

\section{The WISE Data}
\label{sec:data}
\subsection{WISE Survey Strategy}

The WISE satellite followed a 95 minute, Sun-synchronous, polar orbit about 
Earth, always pointing close to zenith and imaging at nearly 90$^{\circ}$
Solar elongation. Over a single orbit, WISE scanned all ecliptic latitudes 
$\beta$ at a particular ecliptic longitude $\lambda$, and then all $\beta$ at 
longitude $(\lambda+180^{\circ})$. Successive  8.8~s, 0.8$^{\circ}$ frames 
overlap by $\sim$10\% in the scan direction, while corresponding frames in 
successive orbits have $\sim$90\% overlap.


Because the All-Sky Release spans just under seven months, the vast majority
of the sky at low ecliptic latitude has a resulting integer coverage of 10-12.
For our purpose of mapping diffuse emission, this substantial redundancy
is useful in compensating for single-frame artifacts such as cosmic rays and 
satellite streaks. However, for typical locations on the sky, all 10-12 WISE 
epochs occur within a single $\sim$20 hour time period. Unfortunately, 
the Moon and bright Solar System objects do not move far enough on this
timescale to be rejected.


\subsection{WISE Data Products}
All WISE data products utilized throughout this work, including 
single exposures and catalogs,  are drawn from the All-Sky Data Release 
\citep{cutri12}. The All-Sky Release features two distinct options for 
accessing WISE images. The ``Atlas images'' are a set of 18,240 co-adds, 
1.5$^{\circ}$ on a side. While it might seem convenient for these co-adds to 
form the basis of our 12$\mu$m dust map, various considerations led us to 
conclude that we were best served by starting our analysis at the 
single-exposure level, then assembling our own custom stacks. First, each 
Atlas co-add is designed only to minimize the appearance of exposure 
boundaries within its footprint, but will not necessarily agree with its 
neighboring Atlas tiles on their overlaps. Additionally, the Atlas images 
contain all compact sources observed by WISE, which require extremely careful 
PSF subtraction in order to isolate the diffuse cirrus. The Atlas PSF must be 
defined in a complex way and would be difficult both to model and apply, given 
the multiple epochs combined at each Atlas pixel and the fact that the Atlas 
images have been smoothed.

Because of these Atlas co-add considerations, we have instead obtained WISE 
imaging data by downloading all $\sim$1.5 million $W3$ exposure sets, 
including the ``intensity'', ``mask'',  and ``uncertainty'' files. The single 
exposures are referred to as ``Level 1b'' (L1b) images, are archived by the 
NEOWISE project \citep{mainzer11}, and are the lowest-level publicly available 
WISE imaging data. L1b images are square, 1016 pixels on a side, and have 
a platescale of 2.75$''$/pixel. The $x$ ($y$) axis of each L1b exposure is 
oriented perpendicular (antiparallel) to the scan direction. 

To remove compact sources, we also make use of the WISE All-sky Source 
Catalog ($\S$\ref{sec:ptsrc}) and WISE All-Sky Known Solar System Object 
Possible Association List ($\S$\ref{sec:sso}).

\subsection{L1b Metadata Cuts}
\label{sec:meta}
We discard a small subset of L1b exposures which, based on metadata alone, 
we know in advance will not be worthy of further processing. Our cuts generally
follow the recommendations and cautions of \cite{cutri12}. Specifically, we 
reject frames with \verb|saa_sep| $\le -5^{\circ}$, or \verb|moon_sep| 
$\le 12^{\circ}$, or \verb|w3msknumsat| $>$ 500,000 pixels, or with 
\verb|dtanneal| $\le$ 1,000~s at high ecliptic latitude. These cuts remove 
4.9\% of the L1b images.

\section{Point Sources}
\label{sec:ptsrc}

A major goal of our L1b processing is the PSF subtraction of each
pointlike source from every exposure in which it appears, along with correction
of the artifacts associated with bright sources. In order to model the PSF and 
artifacts, and eventually to subtract appropriately centered/rescaled PSFs, we 
require a point source catalog. We opted not to build a custom point source 
catalog from scratch. Rather, we make use of the source positions and fluxes 
from the WISE All-Sky Source Catalog, only considering those sources with 
\verb|w3snr| $>$ 2 \citep{cutri12}. 

\begin{figure} 
 \begin{center} 
  \includegraphics[width=3.3in]{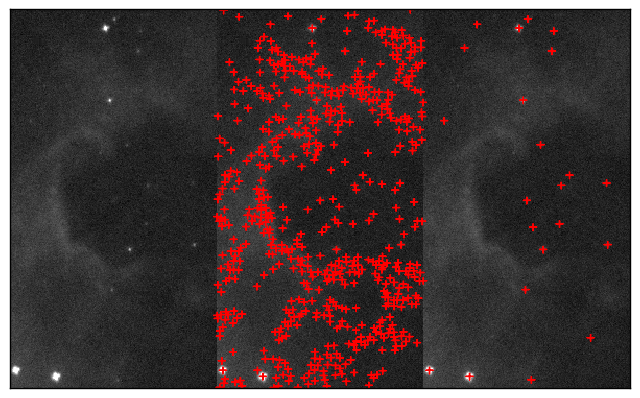}
  \caption{\label{fig:bogus} $0.15^{\circ} \times 0.28^{\circ}$ sub-region 
          of a $W3$ L1b image, centered at ($l$, $b$) = ($212.5^{\circ}$, 
          $-0.45^{\circ}$). Left: L1b data. Center: All-Sky Catalog, all 
          nominal $W3$ detections. Right: list of pointlike sources based 
          on the COG cuts of $\S$\ref{sec:cog}.}
 \end{center} 
\end{figure} 

However, cutting only on \verb|w3snr| $>2$ did not yield a satisfactory 
catalog. In the following subsections, we describe the additional cuts and 
custom augmentations we have made in order to obtain a catalog that best 
subtracts pointlike sources while preserving diffuse nebulosity. In 
particular, we were forced to address issues with very faint and very bright 
sources; at intermediate fluxes, we found the WISE All-Sky Source Catalog 
extremely reliable.

\subsection{Nebula Fragment Rejection}
\label{sec:cog}
The WISE All-sky Source Catalog was engineered to emphasize completeness
over purity. Consequently, a striking feature of the $W3$-detected sample
(\verb|w3snr| $>$ 2) is the tendency for multitudes of dubious, faint sources 
to trace any sufficiently bright, filamentary dust structure (see Figure 
\ref{fig:bogus}). Leaving this issue unaddressed, the filaments in 
PSF-subtracted images take on a distinctly ``chewed up'' appearance, 
compromising our estimate of the dust cirrus.

\begin{figure}[h]
\begin{center}
\includegraphics[width=3.0in]{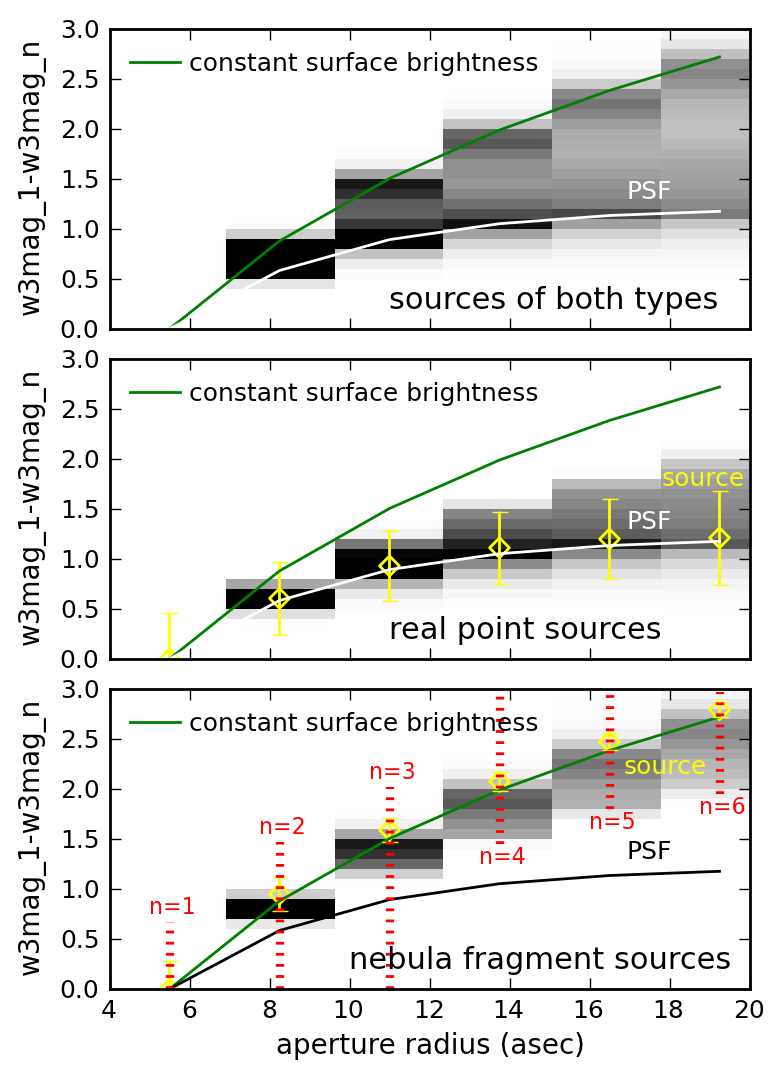}
\caption{Top: $W3$ aperture 1 minus $W3$ aperture $n$ vs. radius for a
 set of $\sim$250,000 All-sky Catalog sources (grayscale), along with the 
expected curve 
for a PSF (white line) and constant surface brightness (green line). Middle: 
same for sources classified as true point sources by our COG cut, with
an example such source shown (yellow). Bottom: same, but for sources 
classified as nebula fragments by our COG cut, along with an example such 
source (yellow).} 
\label{fig:cog}
\end{center}
\end{figure}

We were unable to identify any trivial cuts on e.g. signal-to-noise or 
source color that rectified this problem. However, we observed that the sources
clustered on filaments tend to have a curve of growth (COG) more 
similar to that of a constant surface brightness profile than that of a PSF 
(see Figure \ref{fig:cog}). We measured the COG via a series of aperture
magnitudes, \verb|w3mag_n|, for \verb|n| = $1, \ldots, 6$,  available in the 
All-Sky Catalog. The PSF COG was computed by integrating our model from 
$\S$\ref{sec:psf} in the relevant apertures. We were able to satisfactorily 
reject nebula fragments by cutting those sources with COG more consistent with 
constant surface brightness than with the $W3$ PSF model. We also discarded 
sources with an undefined error \verb|w3sigm_1| on their innermost aperture 
magnitude, \verb|w3mag_1|. The PSF versus constant surface brightness 
comparison was calculated in the sense of minimum absolute deviation, as the 
majority of All-Sky Catalog errorbars (\verb|w3sigm_n|) were undefined for the 
faint sources of interest, precluding a formal likelihood analysis.

The objects thus removed constituted a non-negligible fraction of All-Sky 
Catalog sources with \verb|w3snr| $>$ 2. Prior to the COG cut, the All-Sky 
Catalog contained $\sim$167 million sources detected in $W3$; applying the 
COG filter reduced the number of remaining $W3$ sources to $\sim$41 million.

\subsection{Custom Bright Source Fluxes}
\label{sec:fluxes}
By employing our detailed model of the PSF wings (see $\S$\ref{sec:psf}), we 
were able to derive custom bright source parameters that yielded cosmetically 
optimal PSF subtractions. Our procedure ignores the saturated core of each 
bright source, fitting only an annulus containing the wings.

Specifically, we fit custom fluxes for those $\sim$$15$k sources in the 
All-Sky Source Catalog brighter than \verb|w3mpro| = 2  ($>$5~Jy). This 
threshold has no special significance with regard to the WISE instrumentation, 
and its choice is driven by pragmatic concerns relating to e.g. our available 
computational resources. To fit each bright source, we consider an annulus 
about its centroid with outer radius 2.1$'$. The inner boundary of the annulus 
was determined via a flood-fill of pixels near saturation ($>$$20,000$~DN), 
starting from the centermost pixel. After dilating this saturation mask by 3 
L1b pixels, we excluded the resulting interior region from the flux fit.  We 
then perform a linear least squares fit to the annulus pixel values. The model 
is the sum of a rescaled PSF and a constant offset representing the background 
level. Thus, there are two parameters: the background offset and PSF rescaling 
factor, from which we infer the source flux. For each source, such a fit is 
performed for each L1b image in which the entire fitting annulus falls  within 
the image boundaries. The L1b pixels of each fit are inverse variance weighted 
based on the corresponding  L1b uncertainty image. The flux assigned is taken 
to be the median of the individual L1b fluxes. Typically 9 L1b fits per source 
are performed.

We also computed custom centroids for bright sources in order to achieve 
cosmetically optimal PSF subtractions. After computing our custom fluxes, we 
recompute the centroid by fixing the background and flux, but fitting $x$, $y$ 
offsets via Levenberg-Marquardt chi-squared minimization.

\begin{figure*} 
 \begin{center} 
  \includegraphics[width=7.0in]{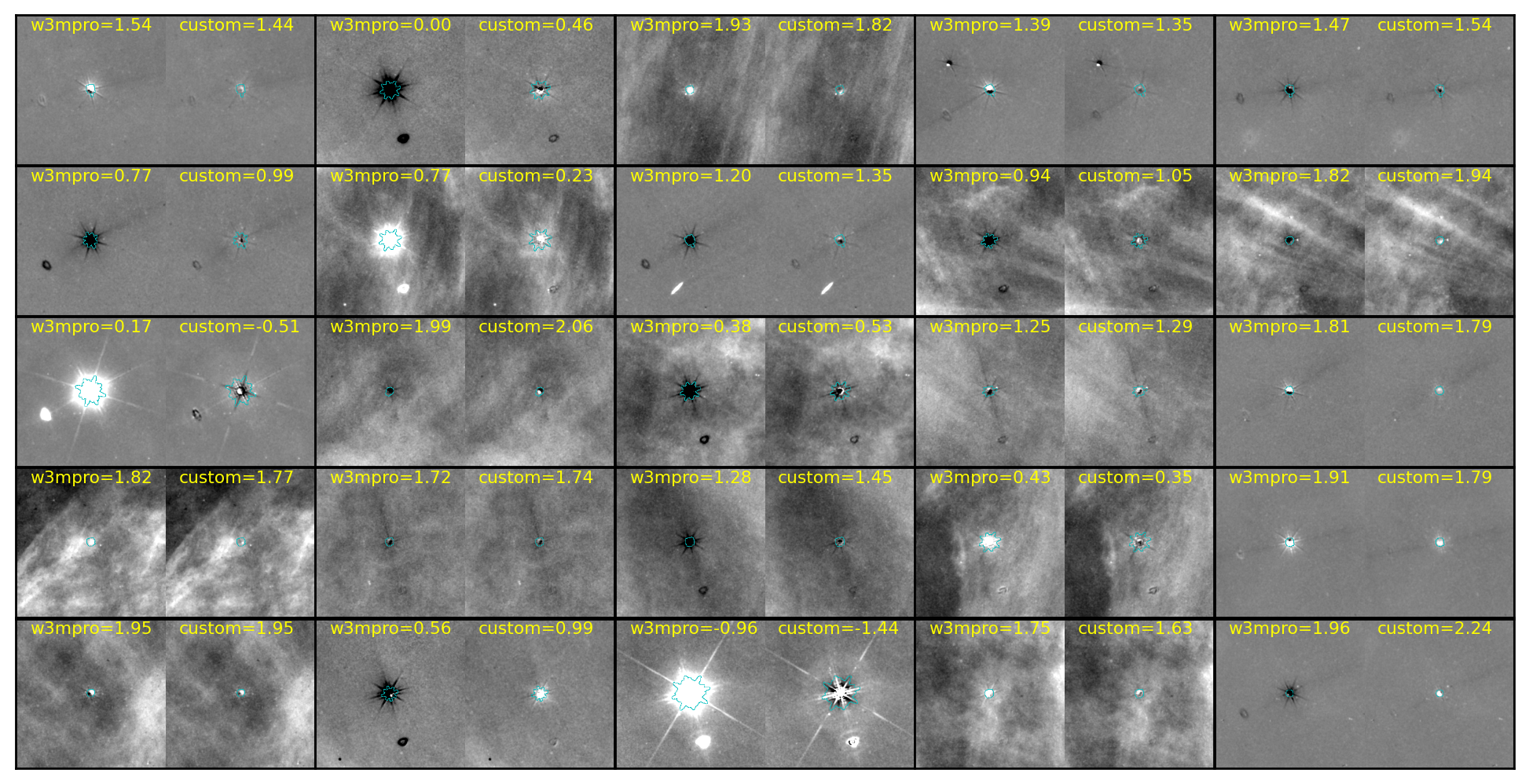}
  \caption{\label{fig:resid} Comparison of bright source subtraction 
residuals using All-sky Source Catalog parameters versus our custom-fitted 
parameters. We have optimized the fit to subtract the PSF wings. Blue outlines
indicate pixels flagged with our ``bright region of point source profile''
mask bit (see Table \ref{table:mask}). The diffraction spikes and ghost also 
appear explicitly in the bit-mask.}
 \end{center} 
\end{figure*}

Figure \ref{fig:resid} shows a representative sample of 25 bright sources 
drawn from $\sim$$300$ square degrees of the sky which we processed in turn 
with the raw All-sky Catalog parameters and our custom parameters. On the
whole, our custom parameters clearly provide substantial 
improvement in the monopole and dipole components of the residuals. For the 
purposes of this work, we only care about the cosmetic quality of our point 
source subtracted maps, and have therefore chosen bright source subtractions 
based on our custom parameters. We caution, however, that our custom bright
source parameters have not been thoroughly vetted against other mid-infrared 
calibrators, whereas those of the official All-sky Source Catalog have been
\citep{cutri12}.

\section{Point Spread Function}
\label{sec:psf}
 
\begin{figure*} 
 \begin{center} 
  \includegraphics[width=7.0in]{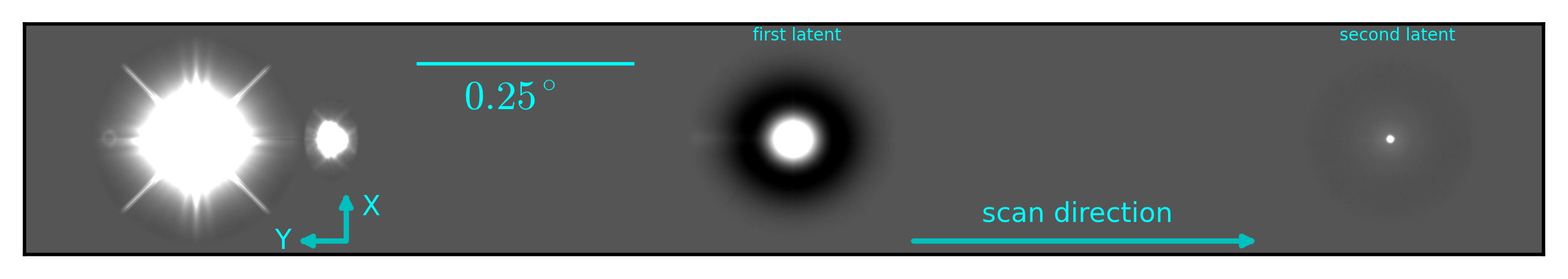}
  \caption{\label{fig:imprint} Our model for the imprint of a single bright 
                          point source, including PSF wings, core, ghost 
                          and first, second latents. }
 \end{center} 
\end{figure*}

Because the space-based WISE mission did not have to contend with atmospheric 
effects, we expect its PSF to be rather stable, and that a detailed 
characterization will lead to excellent PSF subtractions. Figure 
\ref{fig:imprint} shows a schematic of our model for the imprint\footnote{We 
use this term to refer to the PSF and additional non-linear after-effects in
subsequent exposures.} 
of a 
single bright point source in the $W3$ data. The model includes some features
common to astronomical imaging: a bright PSF core, faint but extensive wings, 
and diffraction spikes. Additionally, appearing within the same exposure as 
the bright source itself, a doughnut-like optical ``ghost'' results from the 
WISE optics. This ghost is offset 9.5$'$ in the scan direction relative to the 
source centroid. When a source is bright enough to saturate in its core, the 
following exposure one frame cycle later displays a saucer-like 
electronic defect termed the ``first latent''. Another frame cycle after
 that (two exposures following the bright source detection) the electronics 
are still recovering and a fainter ``second latent'' is observed. For further 
details about these effects, see the Explanatory Supplement \citep{cutri12}. 
In the following subsections we describe our models for each of these 
features, which we later use to PSF subtract pointlike objects and correct 
bright source artifacts.

Our PSF modeling was iterative; after creating an initial model for the PSF 
wings based on out-of-the-box fluxes from the All-Sky Catalog, we used this 
model to fit custom bright source fluxes ($\S$\ref{sec:fluxes}). We then fed 
our refined fluxes back into the PSF/artifact modeling procedure. The models 
eventually used to create the stacks of $\S$\ref{sec:tiles} were all based on 
our custom bright source fluxes. Henceforth, the name \verb|w3mag| will refer 
to our custom bright source fluxes, whereas \verb|w3mpro| will refer 
specifically to values quoted directly from the All-Sky Catalog.

\subsection{PSF Core}
\label{sec:psf_core}
We cannot rely upon one single sample of WISE sources to model all portions of 
the $W3$ PSF. For instance, the extremely bright sources that have high 
signal-to-noise in their outer wings will be saturated in their cores, and thus
useless in modeling the innermost several FWHM.

$W3$ sources begin to saturate in their very central pixels at 3.5 $<$
\verb|w3mpro| $<$ 4.0. In order to leverage the brightest non-saturated 
sources, we modeled the PSF's inner 1.15$'$ (10.6 FWHM, the ``core'') 
using $\sim$4600 All-Sky Catalog sources with 4 $<$ \verb|w3mpro| $<$ 5 and 
$|b|$ $>$ $40^{\circ}$.


To fully reap the benefits of PSF subtraction at the single exposure
 level, our model must include spatial PSF variation across the chip, as
a function of source centroid ($x_c$, $y_c$). We begin by gathering cutouts 
from individual L1b images in which each source in our sample appears. These 
cutouts include those in which the PSF core is only partially 
contained within the image. This permits us to model the PSF variation all the 
way to the chip's edges. Typically 10 such cutouts per source were collected.
We sinc shift each cutout in both $x$ and $y$ directions by a fraction 
of a pixel such that all cutouts are precisely centered on the source 
centroid. Additionally, each L1b cutout is background subtracted and rescaled 
to an arbitrary common source flux. The ($x_c$, $y_c$) location of each 
cutout's centroid within its original L1b image is also stored.

For each pixel location with respect to the PSF model centroid, the set 
of rescaled pixel values available from these cutouts is fit with a 
third order polynomial surface in ($x_c$, $y_c$). We chose third order because 
our measurements of the PSF FWHM as a function of 
position in the chip revealed a dependence that required third order terms. The
per-pixel fit is performed independent of other neighboring pixels and carried 
out via weighted linear least squares, including iterative outlier rejection. 
The weights are inverse variances based on propagation of errors from the L1b 
uncertainty masks. The result is a polynomial model of the PSF core wherein 
the zeroth order term represents the PSF at the center of the chip, and higher 
order terms correct for PSF variation across the chip.

\subsection{PSF Wings}
\label{sec:psf_wings}

Our procedure for modeling the PSF wings is quite similar to that employed in 
modeling the PSF core ($\S$\ref{sec:psf_core}). Our model for the PSF wings 
consists of a rectangular region 14.9$'$ on a side, and thus characterizes 
the $W3$ PSF behavior over a diameter of $\sim$140 FWHM. 

In assembling a sample of sources with which to model the PSF wings, we
aimed to use approximately the $\sim$100$^{th}$-1100$^{th}$ brightest 
high-latitude sources on the sky. Specifically, our sample was defined
to be those sources with $-2$ $<$ \verb|w3mag| $<$ 1, and with $|b|$ $>$ 
$15^{\circ}$. This sample contains 1046 unique sources. We found that the 
handful of \verb|w3mag| $<$ $-2$ sources were predisposed towards anomalous 
defects such as bleeding, and thus would contribute systematic problems to 
the PSF model despite having nominally high signal-to-noise in the wings.

\begin{figure} 
\begin{center} 
  \includegraphics[width=3.0in]{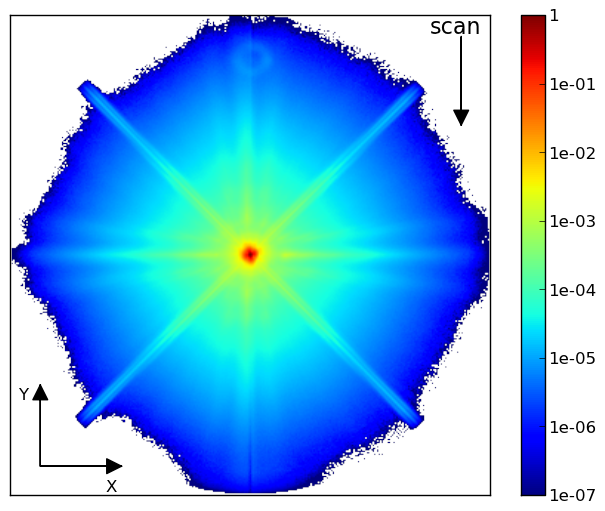}
  \caption{\label{fig:psf} Our model of the $W3$ PSF core and wings at
           chip center, on a logarithmic stretch, tapered according to 
           $\S$\ref{sec:tapering}. The image is 14.9$'$ (325 L1b pixels) on a 
           side. L1b $x$-axis, $y$-axis orientations, as well as the scan 
           direction, are labeled with arrows.}
\end{center} 
\end{figure}

We again collect cutouts as in $\S$\ref{sec:psf_core}, but model the per-pixel 
spatial variation with only first order terms in $x_c$, $y_c$. Higher order 
terms did not appear necessary and tended to destabilize the per-pixel 
polynomial fit in outer regions of the wings where signal-to-noise is low. The 
resulting model of the PSF core and wings, tapered according to 
$\S$\ref{sec:tapering}, is shown on a logarithmic stretch in Figure 
\ref{fig:psf}. The diffraction spikes are not treated specially relative 
to other pixels in the model of the PSF wings.

\subsection{Ghost}
\label{sec:psf_ghost}

The optical ghost location with respect to the PSF core depends strongly on 
the centroid coordinates of its parent source, particularly $x_c$. As $x_c$ 
ranges from one edge of the image to the other, $x_c \approx 0$ to 
$x_c \approx 1015$, the ghost centroid shifts in position relative to $x_c$ by 
0.5$'$ ($\sim$11 L1b pixels). We initially attempted to model this translation 
of the ghost relative to $x_c$ with an approach similar to that in 
$\S$\ref{sec:psf_core}, $\S$\ref{sec:psf_wings}, employing high-order 
per-pixel polynomials. However, we found this model did not satisfactorily 
capture the ghost's translation and became unstable for $x_c$ near the L1b 
edges.

We thus resorted to a ghost model in which the ghost shape remains constant,
but can translate with respect to $x_c$ and can be amplified/suppressed by an 
overall rescaling factor $f_g$ that varies with ($x_c$, $y_c$). To create such 
a model, we again started by extracting many L1b cutouts. The 
same 1046 sources used to model the PSF wings were also used to model the 
ghost. We began by computing the centroid of the ghost in each cutout, and
fit a second-order polynomial model to the ghost's offset relative to 
($x_c$, $y_c$). We then used this ghost translation model to rectify all the 
cutouts to a common fiducial ghost centroid with respect to ($x_c$, $y_c$). 
The rectified cutouts allowed us to compute the ghost profile via a 
weighted average with iterative outlier rejection. To determine 
$f_g$, we modeled each rectified cutout as a sum of an offset and $f_g$ times 
the ghost profile. Finally, we fit a second-order polynomial in ($x_c$, $y_c$) 
to the set of $f_g$ values we obtained. With $f_g = 1$ for ($x_c$, $y_c$) at 
the chip center, our resulting model always has $0.85 < f_g < 1.06$.

\subsection{Latents}

\subsubsection{First Latent}
The first latent always appears with the same centroid coordinates 
($x_c$, $y_c$) as its parent bright source, but one exposure later. By 
``one exposure later'' we do not necessarily mean an increment of one in the 
L1b \verb|frame_num| parameter. Rather, we mean an exposure which follows the 
bright source detection by exactly one frame cycle, 11.1~s. There is not always
such an exposure for every bright source detection. 

We collect the relevant cutouts by identifying those frames that
follow detection of sources with $-3$ $<$ \verb|w3mag| $<$ 1, 
$|b|$ $>$ $15^{\circ}$ by 11.1~s. There are 1088 such sources. Our latent model
is built on a rectangular footprint, 14.9$'$ on a side. To characterize the 
first latent shape, we assume the latent scales linearly with source flux, as 
do other PSF components. We then rescale our cutouts, and determine the latent 
profile via a weighted average with iterative outlier rejection.

In reality, the first latent does not scale linearly with parent source flux. 
As parent flux decreases, the first latent amplitude increases relative to the 
parent source flux (until the parent source no longer saturates, in which case 
no latent will arise). Our model accounts for this fact by assuming the latent 
shape to be independent of parent flux, but allowing for a scalar nonlinearity 
factor $f_l$ that is a function of parent flux. $f_l$ is defined to be a 
correction relative to perfect linear scaling, and would be identically unity 
if the latent behaved as a true component of the PSF, proportional in amplitude
to parent source flux. We compute $f_l$ by creating a first latent model in 
each bin of 0.2 mag and computing their relative amplitudes. A smooth model 
for $f_l$ results from fitting a polynomial to these relative amplitudes. $f_l$
varies between 0.38 and 1.25.

\subsubsection{Second Latent}
\label{sec:mlat2}

We similarly take the second latent to be an artifact appearing 22.2s after 
a bright source detection, with ($x_c$, $y_c$) identical to that of the parent 
source. The procedure for fitting the second latent profile is analogous to 
that for the first latent. To model the second latent we use those sources with
$-3$ $<$ \verb|w3mag| $<$ $-0.5$ and $|b|$ $>$ $15^{\circ}$. Our second latent 
model is rectangular, 12.4$'$ on a side. Because the second latent appeared 
less strongly nonlinear and was more difficult to characterize in narrow bins 
of \verb|w3mag| than the first latent, no nonlinearity correction was 
computed. We also found that higher-order latents were too faint to reliably 
model.

\subsection{Tapering}
\label{sec:tapering}

Our models for the PSF wings and ghost, as well as the latents, are defined on 
rectangular footprints, and may not necessarily asymptote identically to zero 
precisely at their very edges. Thus, subtracting our rectangular models would 
introduce extra power in modes oriented parallel to the $x$, $y$ axes of the 
chip. For this reason we taper the aforementioned models to zero before 
performing subtractions. This is achieved by premultiplying the models by 
weights that decline smoothly from unity in each model's center to zero at 
each model's boundary. We denote the angular extent of a particular model in 
the $x$ ($y$) direction $\theta_x$ ($\theta_y$). In order to define our
tapering weight we first compute a radial coordinate $r$ relative to the 
model's central pixel:

\begin{equation}
r = \sqrt{(\theta_y/\theta_x)^2 \Delta x^{2} + \Delta y^{2}}
\end{equation}

Where $\Delta x$ and $\Delta y$ are displacements relative to the model 
centroid. For $r \le r_{in}$, the tapering weight is unity, and for $r \ge r_{out}$ the 
tapering weight is zero. For $r_{in} < r < r_{out}$, the weight transitions 
from 1 to 0 linearly with respect to $r$. Table \ref{table:taper} lists 
$\theta_x$, $\theta_y$, $r_{in}$, $r_{out}$ for the models that we taper prior 
to subtraction.

\begin{deluxetable}{ccccc}
\tablewidth{0pc}
\tablecaption{Model Sizes \& Tapering Parameters}
\tablehead{
\colhead{Model} & \colhead{$\theta_x$} & \colhead{$\theta_y$} & \colhead{$r_{in}$} & \colhead{$r_{out}$}
}
\startdata
PSF Wings (bright) & 14.9$'$  & 14.9$'$ &  6.7$'$ & 7.4$'$ \\
PSF Wings (faint)  & 5.3$'$  & 5.3$'$ & 1.9$'$ & 2.6$'$ \\
Ghost          & 6.2$'$  & 4.0$'$  &  1.7$'$  & 2.0$'$ \\
First Latent   & 14.9$'$  & 14.9$'$ &  6.7$'$ & 7.4$'$ \\
Second Latent  & 12.4$'$  & 12.4$'$ &  5.5$'$ & 6.2$'$
\enddata
\label{table:taper}
\end{deluxetable}

\subsection{Relating Counts and Magnitudes}
\label{sec:magzp}

In order to relate magnitudes (\verb|w3mag|) to L1b counts (DN), we must
accurately determine the ``zero-point magnitude'' $m_0$, such that a source 
with integrated flux of 1 count corresponds to \verb|w3mpro| = $m_0$.

We chose to define our $m_0$ with respect to an aperture of diameter
1.8$'$. We then conducted aperture photometry on the same set of high-latitude,
 unsaturated, moderately bright sources used to model the PSF core. In order 
to determine $m_0$, we performed a least-squares fit with iterative outlier
rejection to the trend of \verb|w3mpro| vs. 2.5log$_{10}$($F_{DN}$), $F_{DN}$ 
signifying the total counts within the 1.8$'$ aperture. The result determined 
$m_0 = 17.645$. Reassuringly, the best-fit slope was indeed very nearly unity. 
More generally, we found no evidence for significant nonlinearity of 
\verb|w3mpro| for unsaturated sources at any point throughout this work.


We found that $m_0$ was sufficiently stable with time throughout the 
All-Sky Release portion of the WISE mission to be considered a constant. 
Similarly, we did not find any evidence for temporal variation of the 
PSF shape.

\section{Single Exposure Pipeline}
\label{sec:l1b}

We have built a single-exposure pipeline meant to remove and/or flag
several contaminants affecting the diffuse 12$\mu$m emission we wish to 
isolate. Each of the $\sim$1.5 million L1b exposures is processed 
independently. The pipeline attempts to remove all traces of pointlike sources by combining the All-Sky Catalog (including our custom 
bright fluxes and COG cuts) with our models of the PSF and related defects. SSOs are also 
interpolated over and flagged. Additional single-exposure mask bits are set, 
and propagate into extensions of the tile co-adds, as discussed in 
$\S$\ref{sec:tiles}. The following subsections detail our L1b processing steps.

\subsection{L1b Mask}
\label{sec:msk}
Each raw L1b intensity image is accompanied by a corresponding *-w3-msk-1b.fits
bit-mask provided by the WISE team. We will refer to such files as msk files, 
so as not to confuse them with our own custom L1b masks. The first step in our 
L1b processing is to create a mask of bad pixels in the raw L1b image which we 
must interpolate over, based on the msk file. We dilate the cosmic ray mask
by a kernel of sidelength 3 
pixels. Bad pixels are then taken to be those flagged in the dilated 
CR mask, or with any of the static bad pixel msk bits set, or with a 
NaN value. We immediately interpolate over this mask before proceeding.

\subsection{Static Point Sources}
Next we move on to correcting for static pointlike sources. From our custom 
catalog, we retrieve a list of all those sources whose PSF wings or ghost 
image, as represented by our models, have non-zero overlap with the L1b image 
at hand. This includes objects with centroids that fall outside the boundaries 
of the L1b image.

For each point source in the list, we compute its centroid coordinates 
($x_c$, $y_c$) based on the L1b WCS and create a PSF model including the core, 
wings and ghost. The core and wings are computed from the per-pixel 
polynomials in ($x_c$, $y_c$) described in  
$\S$\ref{sec:psf_core}, $\S$\ref{sec:psf_wings}. The ghost image is derived 
based on the model of $\S$\ref{sec:psf_ghost}, taking ($x_c$, $y_c$) into 
account through the ghost centroid translation and the amplification factor 
$f_g$. When computing these models, we always bound $x_c$, $y_c$ to lie 
within the image, so as to not extrapolate beyond the limitations of our 
various polynomial corrections in $x_c$, $y_c$. We renormalize the PSF 
according to each source's \verb|w3mag| value and the prescription of 
$\S$\ref{sec:magzp}.

Before finally subtracting this PSF model from the L1b image, two further 
steps are taken: (1) we sinc shift appropriately such that the observed and 
model source centroids match at the sub-pixel level (2) we  apply the tapering 
prescription of $\S$\ref{sec:tapering} so as not to introduce rectangular 
boundaries into the model-subtracted images. These two steps are also applied 
before subtracting the first and second latent models as described in 
$\S$\ref{sec:lat1}, $\S$\ref{sec:lat2} below.

It was not necessary to apply the full 14.9$'$ model of the PSF wings to 
arbitrarily faint sources. Rather, for sources deemed faint, we subtracted a 
5.3$'$ cutout of the PSF core/wings. This smaller model does not include the 
ghost. We set the threshold distinguishing bright versus faint at \verb|w3mag|
 $ = 4.2$ for typical coverage, though the exact cutoff was scaled on a 
source-by-source basis to account for coverage \verb|w3m|.

For sufficiently bright sources, the innermost pixels of the PSF subtraction
tended to show strong residuals. Consequently, we interpolate over such 
regions, using a circular mask of size that varies with \verb|w3mag|. Such 
interpolation only occurs for sources with \verb|w3mpro| $\le$ 9. We flagged 
pixels interpolated over in this manner with bit 4 of our mask. These
interpolations affect a very small fraction, $<$0.25\%, of the sky.

Based on the appropriately shaped and scaled PSF model derived for each source,
a corresponding bit-mask is created. Saturated pixels within the PSF core 
($>$30,000~DN) are flagged with bit 0. Bright regions of the PSF profile 
($>$250~DN) are flagged with bit 2. Pixels within the ghost model that have 
values greater than 10$^{-4}$ times the peak value of the PSF core are flagged 
with bit 1. This mask bit is meant to trace the general outline of the entire 
ghost model. Especially bright pixels within the ghost model ($>$20~DN) are 
flagged with bit 5.

\subsubsection{First Latent}
\label{sec:lat1}
The presence of a first latent in a given L1b frame indicates
that a bright source at the latent's ($x_c$, $y_c$) location was observed in a 
frame 11.1~s prior. For each L1b frame, we attempt to identify the frame 
11.1~s prior to it. If no such frame exists, then no first latent corrections 
are made. If such a frame exists, we obtain a list of point sources bright 
enough to induce a first latent that were observed within the prior frame. The 
threshold for inducing a latent is \verb|w3mag| $<$ 4. An appropriately 
rescaled latent model, taking into account the nonlinearity correction $f_l$, 
is then subtracted at the location of each bright source in the prior exposure.
The first latent flag (bit 3, see Table \ref{table:mask}) is set for pixels 
within a diameter of 9 pixels relative to each first latent centroid. 

\begin{deluxetable}{cl}
\tablewidth{0pt}
\tablecaption{Bit-mask summary}
\tablehead{
\colhead{Bit} & \colhead{Description}
}
\startdata
0 & saturated point source core \\
1 & point source ghost \\
2 & bright region of point source profile  \\
3 & first latent of point source \\
4 & PSF subtraction residual interpolated over  \\
5 & bright region of point source ghost \\
6 & SSO interpolation \\
7 & resolved compact source \\
8 & second latent of point source \\
9 & third latent of point source \\
10 & fourth latent of point source \\
11 & bright SSO ghost \\
12 & bright SSO latent \\
13 & point source diffraction spike \\
14 & saturated pixel not in static mask \\
15 & Moon contamination \\
16 & RC3 optical galaxy \\
17 & big object (M31, LMC, SMC) \\
18 & Solar System planet \\
19 & reference comparison failure \\
20 & line-like defect \\
21 & low integer frame coverage \\
22 & ecliptic plane
\enddata
\label{table:mask}
\end{deluxetable}

\subsubsection{Second Latent}
\label{sec:lat2}
The procedure for correcting and flagging the second latent is virtually
identical to that for correcting the first latent. The important distinction
is that the relevant list of bright sources will correspond to the frame 
22.2~s prior to the L1b image being processed, if such a prior frame exists. 
We set the threshold for second latent correction at \verb|w3mag| $<$ $-0.5$.

\subsubsection{Third, Fourth Latents}
\label{sec:lat34}
As mentioned in $\S$\ref{sec:mlat2}, we only modeled the first and second 
latents. While we did not model the third and fourth latents, we do include 
mask bits to flag their locations (see Table \ref{table:mask}). We carefully 
inspected the imprints of very bright sources in early versions of our stacks 
to determine which sources required mask bits to be set for third and
fourth latents. We found that sources with \verb|w3mag| $<$ 0 required the 
third latent to be flagged, and those with \verb|w3mag| $<$ $-1.5$ required 
the fourth latent be flagged as well. Flagging the third (fourth) latent simply
involved acquiring the source list for the frame 33.3~s (44.4~s) prior to the 
frame under consideration.

\subsubsection{Diffraction Spike Mask}
\label{sec:spk}
While the diffraction spikes were modeled simply as a part of the PSF wings, we
judged that a special mask bit labeling potential diffraction spike residuals 
might be of value. Using our PSF model, we calibrated the angular extent over 
which diffraction spikes would be detected as a function of source flux and 
integer coverage, conservatively assuming all epochs to have identical scan 
direction and negligible background structure. The threshold for assigning a 
diffraction spike mask is \verb|w3mag| $<$ 4.25 for typical coverage.

We then computed two corrections to this simplistic model to handle relevant 
edge cases. The first problematic scenario arises near the ecliptic poles, 
where scans converge from many different directions and integer coverage is 
very high. Applying the naive model would result in excessively large, nearly 
circular diffraction spike masks in our tile stacks at very high $|\beta|$. 
Consequently, we create an effective magnitude by adding to \verb|w3mag| a 
term that accounts for diffraction spike attenuation due to varying L1b 
exposure orientation. We then use this effective magnitude instead of 
\verb|w3mag| to compute the appropriate diffraction spike extent.

The second case is that of bright, highly structured background emission, e.g. 
in the Galactic plane. In such regions, the systematic variation and elevated 
Poisson noise of the background will diminish the angular extent over which 
diffraction spikes and their residuals are readily apparent. To quantify such 
background structure, we rely upon SFD \IRAS~100$\mu$m (i100). We therefore 
computed $\mathcal{R}_{100}$, an $N_{side} = 64$  HEALPix map 
\citep{gorski05}, wherein each $\sim$$1^{\circ}$ pixel stores the RMS of the 
i100 values sampled at the $N_{side} = 512$ pixel centers nested within its 
boundaries. $\mathcal{R}_{100}$ is typically of order $10^{-1}$ MJy/sr at 
high $|b|$, but can be of order 10$^{3}$ MJy/sr in bright regions like the 
Galactic center. A term logarithmic in $\mathcal{R}_{100}$ was added to the 
effective magnitude we used to compute the appropriate diffraction spike mask 
extent for $\mathcal{R}_{100} > 3$ MJy/sr. This correction adequately reduced 
the diffraction spike extent in regions with substantial diffuse structure.

\subsection{Solar System Objects}
\label{sec:sso}
To some extent, the outlier rejection steps in our mosaicking process 
($\S$\ref{sec:tiles}) eliminate moving objects, without any special treatment 
of SSOs in particular. However, there are many cases in which moving objects 
proceed slowly enough so as to contaminate the same location on the celestial 
sphere at multiple epochs. Furthermore, the WISE team provides a convenient 
catalog of potential SSO detections -- the WISE All-Sky Known Solar System 
Object Possible Association List \citep[][henceforth PAL]{cutri12}. For these 
reasons, we chose to incorporate steps for masking and flagging SSOs into our 
L1b processing pipeline.

The PAL uses the orbits of known SSOs to calculate the expected locations of 
these objects within each L1b exposure. If there exists a WISE detection of 
some source nearby to one of these predicted locations, the measured properties
of that source will be recorded in the PAL. Importantly, each PAL entry 
contains an L1b frame identification, coordinates of the source within that L1b
image, and the source's profile-fit magnitude, again designated \verb|w3mpro|. 
In some cases, the associations may be inertial sources, or spurious 
detections, as PAL entries are based on single-epoch photometry. For our 
purposes, sample purity is not of much concern. Even if we interpolate 
unnecessarily in some cases, the redundant WISE coverage will tend to render 
such a mistake imperceptible.

For each SSO in a given L1b exposure, we produce a mask centered about the SSO 
centroid that scales in size with the detected object's brightness. These masks
each contribute to bit 6 of our L1b bit-mask (see Table \ref{table:mask}), 
which denotes the pixels we have interpolated over within the L1b image to
remove the SSO core/wings. The mask about an SSO's centroid consists 
predominantly of a circular region, with diameter varying from 0.3$'$ 
(7 L1b pixels) for \verb|w3mpro| $\ge$ 10 to 3.5$'$ (83 L1b pixels) for 
\verb|w3mpro| $\le$ 1.4. In addition, we apply our diffraction spike model of 
$\S$\ref{sec:spk} to SSOs with \verb|w3mpro| $<$ 3. We interpolate over these 
SSO diffraction spikes at the L1b level, and the affected pixels are also 
flagged with bit 6. We considered PSF subtracting SSOs rather than simply 
interpolating over them, but found that the single-epoch PAL photometry was 
too unreliable and yielded a cosmetically inferior result. 

\begin{figure}[ht]
\begin{center}
\includegraphics[width=3.4in]{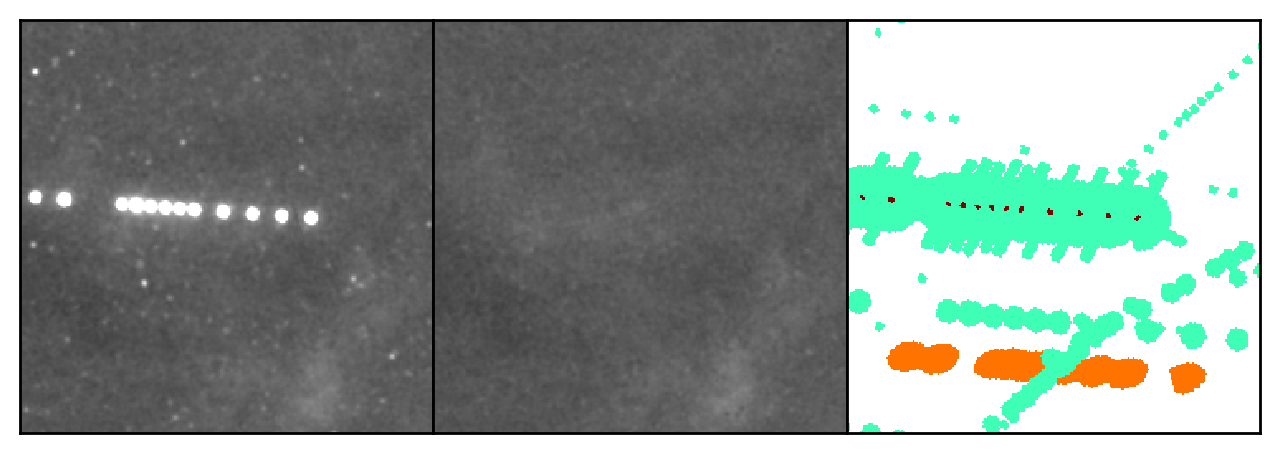}
\caption{\label{fig:sso} $0.4^{\circ} \times 0.4^{\circ}$ co-add
          cutout at ($l$, $b$) = (355.68$^{\circ}$, 43.84$^{\circ}$),
          ($\lambda$, $\beta$) = (226.74$^{\circ}$, 13.19$^{\circ}$),
          illustrating our SSO mask bits. Green, orange, and red represent bits
          6, 11, and 14 respectively (see Table \ref{table:mask}). The bright 
          SSO (381 Myrrha) reveals the detailed nature of bit 6, which 
          incorporates the diffraction spike model of $\S$\ref{sec:spk}.}
\end{center}
\end{figure}

In addition to addressing each SSO's core, wings and diffraction spikes, we
take further actions with regard to artifacts of very bright SSOs. For 
SSOs with \verb|w3mpro| $<$ 2.5, we mask and interpolate over an 
elliptical region surrounding the object's ghost image. SSO ghost 
interpolation is registered in our bit-mask with bit 11. Similarly, for SSOs
brighter than \verb|w3mpro| = 1, we interpolate over the first latent and 
correspondingly set mask bit 12. Figure \ref{fig:sso} provides an example of 
our SSO masking and interpolation results.

\subsection{Bit-mask Summary}
\label{sec:mask_summary}
Table \ref{table:mask} lists all mask bits included in the WSSA tiles of our 
data release. Throughout $\S$\ref{sec:l1b}, bits 0-6, 8-13 have been explained.
Bit 14 flags pixels that are saturated ($>$30,000 DN), but not marked as 
suspect by any of the msk file static mask bits. This can be useful downstream 
in identifying regions of extremely bright nebulosity that have reached 
saturation. Bits 7, 15-18 are only defined at the co-add level and are 
explained in $\S$\ref{sec:mmsk}.

\subsection{Smoothing and Rebinning}

Before writing our processed L1b files, we smooth to 15$''$ FWHM, trim the 
smoothed image to $1000 \times 1000$, and bin down by a factor of two in each 
dimension. We write the outputs of our L1b processing pipeline as 
multi-extension FITS files. For each raw L1b file in the All-Sky Release we 
write a corresponding processed file. The primary extension is the fully 
``cleaned'' image, where we have removed point sources and their artifacts and 
SSOs. Extension 1 is the ``dirty'' image, which is merely a smoothed and 
rebinned version of the raw L1b image after interpolating as described in 
$\S$\ref{sec:msk}. Extension 2 is our bit-mask of $\S$\ref{sec:mask_summary}, 
after dilation and rebinning that mimics the smoothing and rebinning of the 
first two extensions. These processed L1b images are not included in our data
release, but will be made available upon request.

\section{Stacking}
\label{sec:stack}
We chose to co-add our processed L1b images on astrometric footprints 
analogous to those of
the \IRAS~Sky Survey Atlas (ISSA) tiles \citep{wheelock94}. We define a set of
``WSSA'' astrometric footprints with the same central ($\alpha$, $\delta$) 
values and same orientations relative to north as the original ISSA tiles, but 
precessed to J2000. As a result, each of our WSSA tiles does not identically 
cover the same region of the sky as its ISSA counterpart. Also, to accomodate
our 15$''$ FWHM, WSSA tiles 
have a pixel scale of 5.625$''$/pixel, as opposed to 1.5$'$/pixel for ISSA 
tiles. There are 430 such $12.5^{\circ} \times 12.5^{\circ}$ WSSA tiles, with 
centers tracing out a series of isolatitude rings in J2000 equatorial 
coordinates. These tiles are not mutually exclusive; over the full sky, the 
median tile coverage is 1, but the mean coverage is 1.6, with a maximum of 5.

Our basic strategy in producing each WSSA tile is to start from the tile's 
center and build outward, matching each image with the existing stack via a 
single offset before adding it in. Admittedly, this is a simple-minded 
approximate solution to the matrix inversion problem posed by best matching
overlaps between neighboring exposures. In order to ensure that our
simplistic procedure is feasible, we take two additional steps before 
beginning the stacking process, described in the following two subsections.

\subsection{Gradient Correction}
\label{sec:grad}
Overlapping L1b images may differ by more than a simple relative offset. This 
is especially true in regions that were imaged six months apart with opposite 
scan directions or near the ecliptic poles. In such cases, the zodiacal light 
can differ substantially between epochs. Even though the zodiacal light at a 
given epoch generally varies on spatial scales larger than a degree, this 
foreground is so strong that its gradient is readily apparent in single L1b 
exposures. As such, to ensure that our stacking can match neighboring exposures
with a single offset, it is important to compute a correction for each exposure
that removes time-dependent first order structure.

To arrive at such a gradient correction, we compare each L1b exposure to SFD
100$\mu$m on the same footprint. For each cleaned L1b exposure, we interpolate 
generously over bright source artifacts using our masks, then smooth to $6.1'$ 
FWHM to match i100 resolution and bin down 5$\times$ in each dimension. We 
denote each pixel with coordinate $x_i$, $y_i$ in this smoothed, rebinned $W3$ 
image as $w_i$. We then fit the following model to $w_i$:

\begin{equation}
w_i = f_{wi}i_i + \nabla_x x_i + \nabla_y y_i + C 
\end{equation}

Where $f_{wi}$, $\nabla_x$, $\nabla_y$, and $C$ are the scalar parameters to 
be fit and $i_i$ are the i100 values corresponding to $w_i$. The fit is 
performed via unweighted linear least squares, with iterative outlier 
rejection. $\nabla_x$, $\nabla_y$ then specify $x$, $y$ gradients that make 
the processed L1b exposure most consistent with Galactic emission, free of 
slowly varying time-dependent foregrounds. $f_{wi}$ is a factor that allows 
small-scale dust structure to be nulled, accounting for the fact that 
conversion factor between $W3$ and i100 will be different for different L1b
pointings. Similarly, $C$ absorbs any constant offset arising from the WISE and
i100 zero points. Our model for the time-dependent gradient in each exposure is
computed from $\nabla_x$, $\nabla_y$, and subtracted before the mosaicking 
process begins. We only desire the gradient terms, and make no use of $f_{wi}$,
$C$ downstream.

\subsection{Pairwise Rejection}

Despite our metadata cuts ($\S$\ref{sec:meta}), some L1b images with severe 
systematic problems remain and could potentially corrupt the stacking process 
if allowed to contribute to our co-adds. To remove such images, we employ a 
pairwise comparison between overlapping L1b exposures. 

Specifically, each gradient-corrected, processed L1b image is compared to up to
20 of its nearest neighbors at the locations of $N_{side}$ = 2048 HEALPix pixel
centers on which the frames overlap. For each pair, we compute the RMS of the 
differences between the sampled values. Then, for each image, we compute the 
median of these RMS values, $\mathcal{R}_{med}$. We then reject the highest 
$\mathcal{R}_{med}$ frame with $\mathcal{R}_{med} > 500$ DN, updating 
$\mathcal{R}_{med}$ for all other frames after each rejection, until no frames 
remain with $\mathcal{R}_{med} > 500$ DN. $\sim$0.4\% of images are rejected 
by the pairwise comparison, but these frames would have inflicted a 
disproportionate, negative influence on our stacks given the severity 
of their systematic problems.

\subsection{Initial Stack}
\label{sec:init}
For each WSSA tile, we co-add the constituent L1b exposures onto the relevant
astrometric footprint, with pixel scale of 5.625$''$/pixel. 
12.5$^{\circ} \times$12.5$^{\circ}$  co-adds are thus 8,000 pixels on a side. 
Such tiles are well-sampled because we have smoothed our processed L1b 
images to 15$''$ FWHM.

We obtain a list of L1b exposures that will overlap the tile's astrometric
footprint, sorted by proximity to the tile center. We apply the gradients of 
section $\S$\ref{sec:grad} to each cleaned L1b image, and then use the pairwise
 rejection step to reject the typically small number of problematic exposures. 
Starting with the exposure closest to the center, we add the cleaned L1b 
exposures into the stack one at a time, applying a single offset to make each 
exosure best agree with the existing stack. This offset is simply the median 
value of the difference between the exposure at hand and the existing stack on 
their overlap. As we stack, we also keep track of the minimum and maximum 
single-exposure pixel values at each pixel in the co-add footprint. We also 
store an integer coverage map, incrementing the relevant pixels each time an 
L1b image is added to the stack. For locations in the completed co-add that 
have coverage $>$2, we subtract out the average of the minimum and maximum 
images at those locations, and adjust the coverage map appropriately. We will 
refer to the average of the minimum and maximum images as the 
``transient artifact image''. This outlier rejection step eliminates satellite 
streaks and any other unusual single-epoch features not specifically addressed 
by our L1b processing pipeline. We also apply the per-frame offsets to the 
dirty extension of our L1b images to create a corresponding stack with all 
point sources, bright source defects, and SSOs present.

\subsection{Reference Comparison}

With our initial WSSA tile stacks in hand, we can perform some further 
refinements, using the existing stack as a reference. There are two typically 
low-level problems we address by making use of our reference stack: (1) 
per-quadrant polynomial offsets in individual L1b exposures and (2) striping 
artifacts in L1b images.

\subsubsection{Per-quadrant Polynomial Warps}
\label{sec:warp}
We have already addressed scalar intensity offsets and gradient corrections on 
a per-exposure basis. But within a given exposure, the four quadrants can be 
offset relative to each other due to four-amplifier detector readout. These 
offsets can be simple scalars, but in some cases involve higher order 
polynomials (e.g. a gradient across one quadrant that is absent in the other 
quadrants). We refer to these polynomial offsets as per-quadrant ``warps''. For
fixed position on the celestial sphere 
within our co-adds, such warps in the various contributing L1b quadrants are 
not correlated. As such, any warping will be suppressed by $\gtrsim$90\% in 
our initial stacks.

Thus, we may reasonably consider our initial stacks to be nearly free from 
the effect of such per-quadrant warping, and as a result use them as reference 
templates to correct individual L1b quadrants. Doing so can help eliminate
any jagged L1b quadrant edges that persist at low levels in the initial 
stacks. For each tile, we gather the same sample of L1b exposures as in 
$\S$\ref{sec:init}. We then rebuild the mosaic from scratch, one L1b quadrant 
at a time, first fitting a per-quadrant warp relative to the reference, and 
then subtracting the warp model before adding the corrected quadrant into the 
revised stack. 

\begin{figure}
\begin{center}
\includegraphics[width=3.2in]{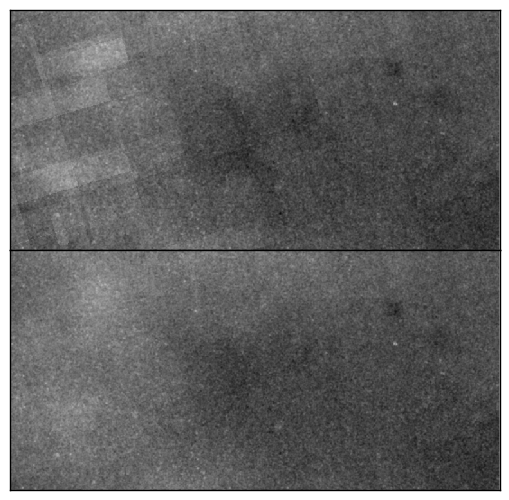}
\caption{\label{fig:warps} $1.2^{\circ} \times 0.6^{\circ}$ region within WSSA 
                      tile 221, at $(l$, $b) \approx (5^{\circ}$, 
                      $48^{\circ}$). Top: before reference warping. Bottom: 
                      after reference warping, with jagged L1b exposure 
                      boundaries eliminated.}
\end{center}
\end{figure}

Our warp model is simply a polynomial in the the relevant cleaned L1b 
quadrant's $x$, $y$ coordinates. For each such quadrant, we fit a fourth order 
polynomial to the difference between the cleaned L1b image and the reference 
stack. The fit is performed via unweighted linear least squares. With the 
best-fit polynomial computed, we evaluate the $\chi^2$ goodness of fit between 
the reference stack and the corrected L1b quadrant. We take the expected pixel 
variance in each pixel of the cleaned L1b image to be:

\begin{equation}
\sigma_i^2 = f_{s}^2(R^2 + \frac{N_{i}}{g}) + (\delta g N_{i})^2 + (\nabla N|_{i}  \delta_{\theta})^2
\end{equation}

Where $N_i$ is the cleaned L1b pixel value in DN, $R$ = 16.94 DN is the 
readnoise, $g$ = 6.83 $e^{-}/$DN is the gain, $f_s$ is a constant that
accounts for our smoothing/rebinning, $\delta g$ is an estimate of the
fractional gain variation,  $\nabla N|_{i}$ is the magnitude of the reference 
image gradient evaluated at pixel $i$, and $\delta_{\theta}$ is an estimate of 
the L1b astrometric error. $N_i$ is floored at zero in the second term, since 
it is possible that e.g. a bright source residual could have negative values. 
$\nabla N|_{i}$ is calculated based on the reference stack, not the L1b image 
being compared to the stack. Based on studies of point sources presented in 
\cite{cutri12}, we chose $\delta g$ = 0.025 and $\delta_{\theta}$ = 0.4$''$. 
The first two terms account for statistical errors in the cleaned L1b image. 
The second term allows for the possibility that the per-quadrant L1b gain may 
vary slightly. The final term accounts for astrometric imperfections that could
cause the L1b exposure to be out of alignment with the reference stack. We 
assume that statistical errors in the reference stack are negligible relative 
to those in the L1b quadrant of interest.

If the $\chi^2$/D.O.F. is greater than 7 for a particular L1b quadrant after 
applying the best-fit polynomial warp, that quadrant is excluded from the 
stack. 2\% of L1b quadrants were thus rejected. Figure \ref{fig:warps} shows
 a region in which L1b boundaries in the initial stack of 
$\S$\ref{sec:init} are corrected by our quadrant warping procedure.

\begin{figure}
 \begin{center} 
  \includegraphics[width=3.3in]{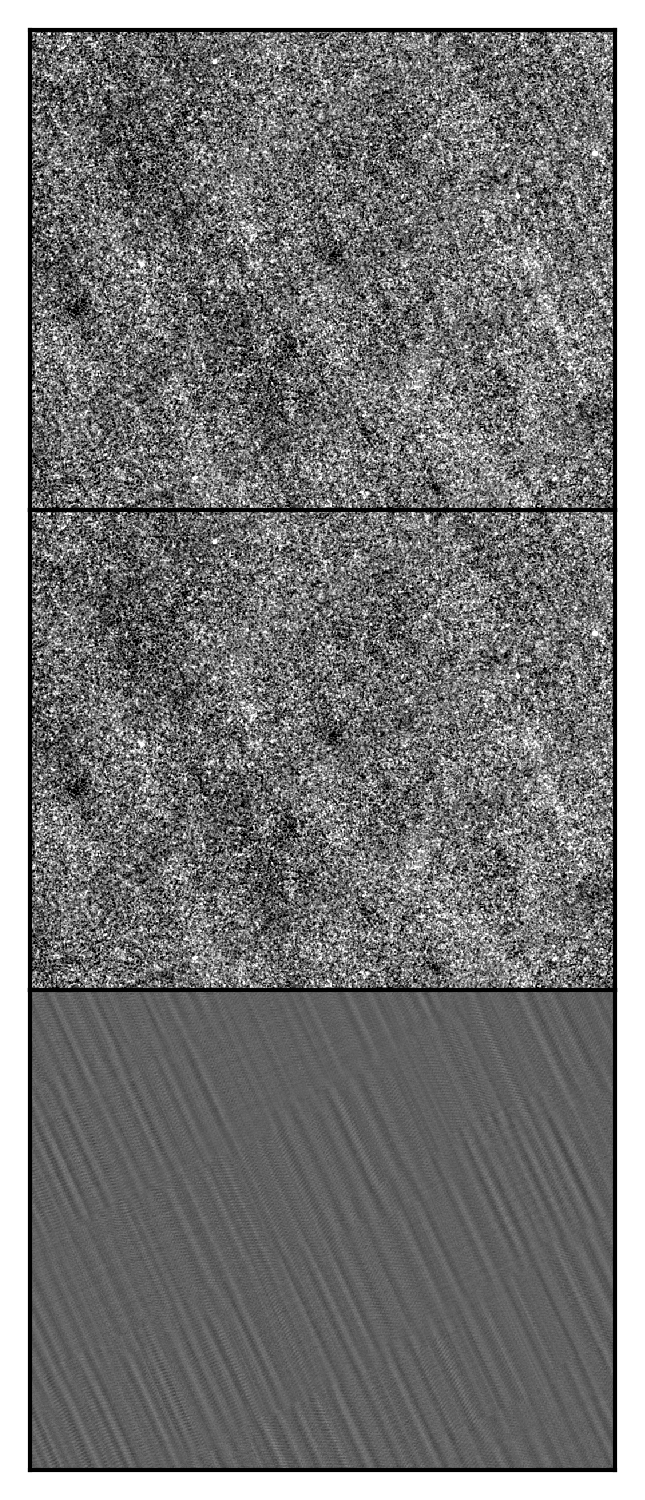} 
  \caption{\label{fig:stripe} $1.6^{\circ} \times 1.3^{\circ}$ sub-region
           within WSSA tile 339, at ($l$, $b$) $\approx$ ($188.7^{\circ}$, 
           $56.2^{\circ}$). Top: before destriping. Middle: after destriping.
           Bottom: stripe model. The stripes are parallel to the scan 
           direction.}
 \end{center} 
\end{figure} 

\subsubsection{Destriping}

L1b exposures display low-level striping. Complicating matters, this striping
occurs with different amplitudes and on different wave numbers from exposure 
to exposure, and even from quadrant to quadrant within single exposures. Still,
a particular set of relatively few modes tend to be enhanced, as the stripes 
within a given quadrant are typically oriented parallel to the scan 
direction and have wavelengths of $\sim$2.5$'$-4$'$. To avoid high-pass
filtering diffuse structure, our destriping never modifies modes larger than 
$4.6'$. We assume the striping to be an additive rather than multiplicative 
defect.

As such, our first step towards correcting L1b striping is to make a mask in 
Fourier space of the typically contaminated modes, using $\sim$20,000 cleaned
L1b images at very high $|b|$, where there are few compact sources and diffuse 
structure is minimal. For each of these L1b exposures, we compute the isotropic
Fourier power spectrum in each quadrant. We label any mode ``contaminated'' 
which has a modulus-squared that is a $>$$5\sigma$ outlier relative to the 
appropriate isotropic value. We construct a mask of frequently contaminated 
modes by counting the number of occurences of such outliers at each pixel 
within the quadrant FFT.

For each quadrant that successfully passes the quadrant warping step of 
$\S$\ref{sec:warp}, we attempt a destriping correction before finally adding
the quadrant into the reference-corrected stack. We take the  FFT of the 
quadrant under consideration and identify any $>$$5\sigma$ outliers relative 
to the isotropic power spectrum. We then modify the FFT values of those 
outliers which coincide with modes we previously flagged as frequently 
contaminated. We rescale such outliers to have a modulus that matches the 
appropriate isotropic value, leaving the phase unchanged. Transforming back to 
the spatial domain yields a destriped image.

For most L1b quadrants, this correction removes all apparent striping 
artifacts. However, there are cases in which the correction can go awry. For 
example, the residuals of bright source artifacts can be aligned with the scan
direction, wreaking havoc with our analysis meant to deal with low-level 
striping of the same orientation. We therefore incorporated destriping into 
our reference comparison process, in order to robustly identify such 
problematic instances. Specifically, we perform a chi-squared comparison 
between each L1b quadrant and the reference stack, both before and after 
destriping. We use the pixel variances of $\S$\ref{sec:warp}. We then apply 
the destriping correction only if (1) the destriping correction improves the 
$\chi^2$/D.O.F and (2) more individual pixels within the quadrant have their 
$\chi^2$ improved than worsened by destriping. Since the striping is most 
apparent in regions of very low signal/background, and the destriping 
correction is very successful in such regions, we find this ``do no harm'' 
approach to be satisfactory. Figure \ref{fig:stripe} shows an example of our 
destriping correction.

\subsection{Mosaic-Level Mask Bits}
\label{sec:mmsk}
The following subsections describe mask bits defined at the mosaic level, 
rather than the L1b level.

\subsubsection{Moon Mask}
\label{sec:moon}
While our metadata cuts remove L1b frames pointed near the Moon 
(\verb|moon_sep| $\le$ 12$^{\circ}$), off-axis Moon-glow can 
adversely affect exposures with much larger \verb|moon_sep| values. In fact, 
prominent off-axis Moon illumination features extend out to many tens of 
degrees away from the Moon's location. Accounting for the WISE survey strategy,
this illumination detrimentally contaminates a non-negligible fraction, 
$\sim$5-10\%, of the sky. Adverse effects on our stacks arise because 
the extended Moon-glow violates our assumption that the differences between 
exposures can be accounted for by an offset and linear gradients. For example, 
Moon-glow diffraction spikes can imprint strong, higher-order 
structure within a single L1b frame (see \citealt{cutri12}, $\S$2.4b).

\begin{figure} 
 \begin{center} 
  \includegraphics[width=3.3in]{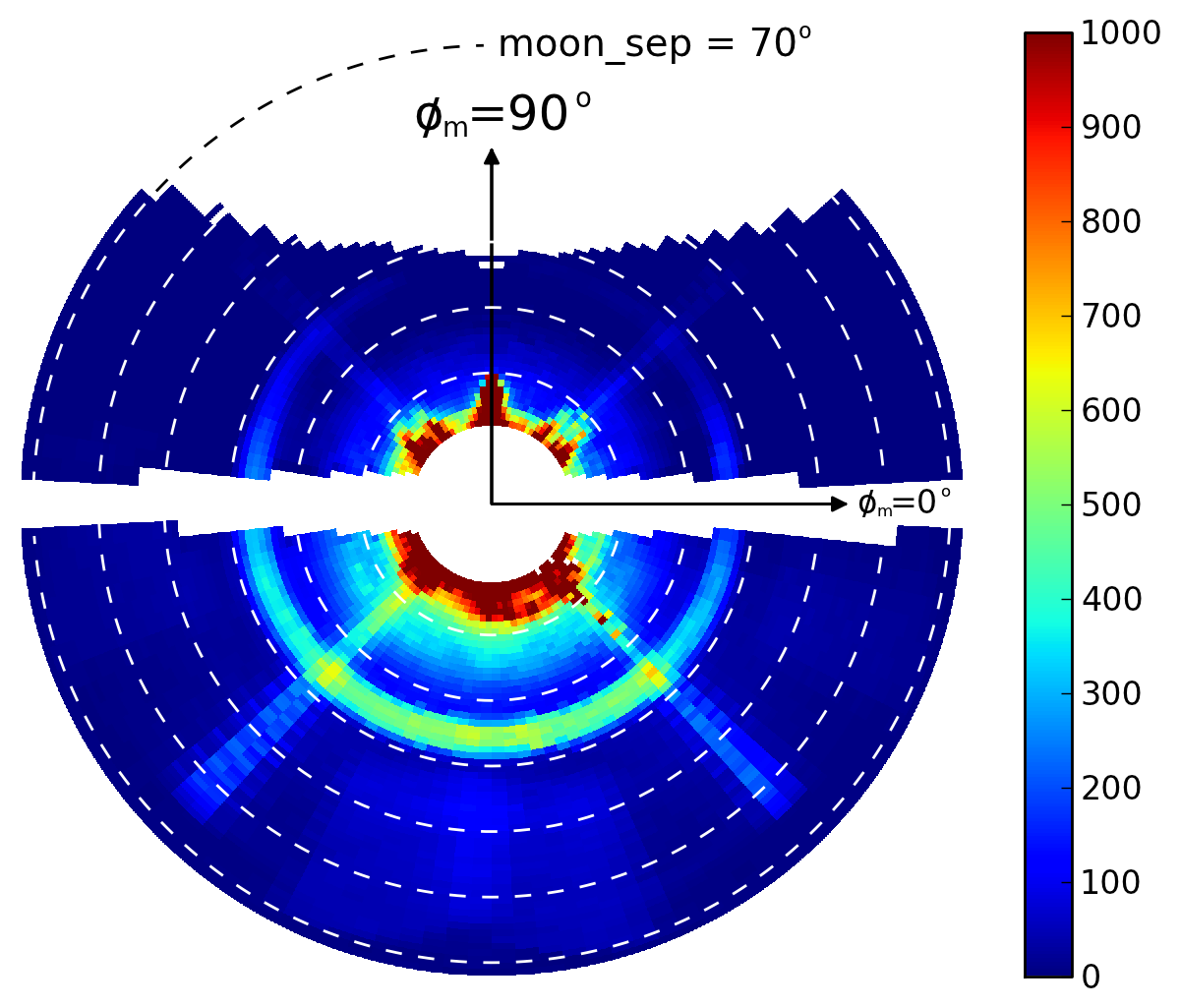}
  \caption{\label{fig:moon} Our model for $W3$ off-axis Moon-glow as a function
                          of angular separation  \texttt{moon\_sep} (radial 
                          coordinate) and position angle $\phi_m$ 
                          (azimuthal coordinate). 
                          The units are $W3$ DN, and dashed white circles 
                          denote intervals of $10^{\circ}$ in 
                          \texttt{moon\_sep}. Diffraction spikes are apparent 
                          out to \texttt{moon\_sep} $>60^{\circ}$. 
                          $0^{\circ}$$<$$\phi_m$$<$180$^{\circ}$
                          tends to correspond to lower Moon phase than 
                          $-180^{\circ}$$<$$\phi_m$$<$0$^{\circ}$, and is thus 
                          fainter.}
 \end{center} 
\end{figure} 

We derived an empirical model of $W3$ off-axis Moon illumination in order to 
flag portions of our stacks which suffer serious Moon-glow corruption. 
In ecliptic coordinates, Moon-contaminated regions trace out a series of 
narrow vertical bands, a few degrees wide and generally separated by 
$\sim$30$^{\circ}$ in $\lambda$. We did not attempt to characterize
 sub-exposure Moon-glow structure, but instead created a model of the additive 
offset to the WISE zero point due to Moon illumination. Our model is not a 
function of ($\lambda$, $\beta$), but rather Moon angular separation 
(\verb|moon_sep|) and Moon position angle relative to the WISE optics 
(termed $\phi_m$ = \verb|moonpa|$-$\verb|w3pa|). To arrive at such a model, we 
generated a full-sky $W3$ map, taking the L1b zero point at face value. We 
then subtracted 1.2$\times$ the Kelsall model to roughly cancel the zodiacal 
light. Next, we smoothed to 6.1$'$, and subtracted scaled SFD i100 assuming 
i100/$W3$ = 17.5, to approximately cancel diffuse Galactic emission. We 
designated particular, narrow ranges of $\lambda$ as Moon-affected, and 
created a Moon-free prediction by interpolating over such regions at constant 
$\beta$. Restricting to $|b|$$>$$15^{\circ}$, the difference between the 
Moon-free prediction and the corresponding L1b data constitutes our Moon-glow 
estimate, which yields the profile shown in Figure \ref{fig:moon} when binned 
in (\verb|moon_sep|, $\phi_m$). We also created analogous profiles in Moon 
phase and Moon distance to enable corrections for these factors when predicting
Moon contamination for individual L1b pointings.

As we stack, at each pixel in the co-add, we keep an integer count of the
number of Moon-affected frames contributing to that pixel. We define 
Moon-affected frames to be those for which our model predicts moon 
contamination of $S_m$$>$100 DN, and $S_m/\mathcal{R}_{100}$$>$450. Pixels with
Moon-contamination count $>$3 at the end of the stacking process are flagged 
with bit 15, as listed in Table \ref{table:mask}.

In rare cases typically associated with moon contamination, all L1b quadrants 
at a particular sky location are rejected during the 
reference comparison step. Mask bit 19 labels such locations, at which we have
tapered in the reference stack to avoid leaving regions of missing data in the 
final mosaic. 0.15\% of the sky is affected.

\subsubsection{Resolved Compact Sources}
Though we have carefully removed pointlike sources, there are many resolved
compact sources in the WISE All-Sky Catalog. These objects, which include 
galaxies and multiple sources, do not belong in our map of diffuse Galactic 
cirrus. Instead of attempting to model and subtract such sources, we have 
simply added mask bit 7 to flag their locations, allowing users to ignore or 
interpolate over the affected regions.

To obtain a catalog of resolved, compact $W3$ sources, we cross-matched WISE
All-Sky Catalog objects with \verb|w3snr| $>$ 2, \verb|ext_flg| $=$ 5 to 
2MASS XSC sources, with a matching radius of 5$''$. We excluded XSC objects
with \verb|cc_flag| = \verb|a| or \verb|cc_flag| = \verb|z|, and additionally
discarded the 3712 XSC sources believed to be fragments of diffuse Milky Way 
dust \citep[][$\S$2.3b]{tmass}. This resulted in a list of $\sim$150,000 
resolved $W3$ sources. In order to obtain the best possible set of attributes 
for each such object, including the semi-major axis to semi-minor axis 
ratio $r_{ab}$ and position angle $\theta_{PA}$, we cross-matched our
resolved $W3$ source list to the PGC catalog \citep{pgc03}, with a matching 
radius of $1'$.

We proceed to merge attributes from the XSC and PGC catalogs. We assign
PGC values for $r_{ab}$ and $\theta_{PA}$ to all PGC-matched sources with
both values available in the PGC catalog. For objects with PGC major axis
$>$1$'$, but lacking either $r_{ab}$ or $\theta_{PA}$ in the PGC catalog,
$r_{ab}$ and $\theta_{PA}$ are assigned to their XSC values. All other
sources are assigned a default value of $r_{ab} = 1$. Relatively large sources 
with PGC major axis $>$1$'$ are assigned their respective PGC major axis 
value. All other sources are assigned a major axis of $1'$. To be conservative 
and avoid overly line-like masked regions, we cap $r_{ab}$ at 2. The LMC, SMC, 
and  M31 are excluded from the bit 7 mask and flagged separately 
(see $\S$\ref{sec:bigobj}).

For each WSSA tile, we then create a resolved object mask by flagging
elliptical regions corresponding to the relevant entries from our 
WISE/XSC/PGC catalog. Over the entire sky, only 0.07\% of pixels are thus 
flagged.

Our cross-matching procedure may occasionally discard true $W3$ sources
 in cases of exotic morphology (e.g. Barnard's Galaxy), or in 
the event of a WISE All-Sky Catalog or XSC failure. To be conservative, we 
created mask bit 16, which uses the available PGC parameters of any prominent
optical galaxies missing from our cross-match to again flag appropriate 
elliptical regions. To isolate such a sample of  `prominent' optical sources, 
we restricted to the subset of PGC sources with an RC3 counterpart \citep{rc3}.
Before generating this mask, we carefully excluded 18 large ($>$5$'$ diameter),
dIrr or dSph galaxies which are completely invisible in $W3$ (e.g. the Fornax 
Dwarf). In all, bit 16 masks only 0.005\% of the sky.


\subsection{Low Integer Frame Coverage}
In regions with low integer frame coverage $<$5, compact sources are listed
only in the WISE All-sky Reject Table, and are excluded from the All-sky Source
Catalog \citep{cutri12}. Rather than attempt to identify valid sources among 
Reject Table entries, we opted instead to flag such regions with mask bit 
21. 1.2\% of the sky is affected.

\subsubsection{Big Objects}
\label{sec:bigobj}
The SFD mask contains a so-called ``big object'' bit, which labels
regions affected by extragalactic sources with exceptionally large
apparent sizes: the LMC, SMC, and M31. Bit 17 of our
masks is meant to flag these same three objects. Our big object mask is 
identical to that of SFD for the irregularly shaped LMC and SMC. For M31, we 
use an ellipse based on PGC catalog parameters.

\subsubsection{Low Ecliptic Latitude}
Unfortunately, certain low ecliptic latitude regions were only observed by 
WISE while obstructed from view by a planet within our Solar System. In such 
cases, the planet appears as an extraordinarly bright, compact source and 
renders measurement of diffuse Galactic emission in its vicinity completely 
hopeless. As a result, we have flagged circular regions about the centroid of 
each planet imprint, typically several degrees in diameter. We
have also added mask bit 22 to flag all pixels immediately within the ecliptic 
plane, as these should generally be handled with extra caution. The approximate
range of ecliptic latitudes flagged is $-2^{\circ}$ $<$ $\beta$ $<$ 
$2^{\circ}$. Additionally, at low ecliptic latitude, we have noticed unexpected
line-like features $\sim$$5.5'$ wide and extending many degrees in length. 
The origin of these lines is not known, and they do not appear to follow 
trajectories obviously indicative of an instrumental or processing malfunction,
e.g. constant $\delta$ or constant $\beta$. Mask bit 20 marks the locations of
such lines, which affect less than 0.15\% of the sky.


\begin{figure}
\begin{center}
\includegraphics[width=3.0in]{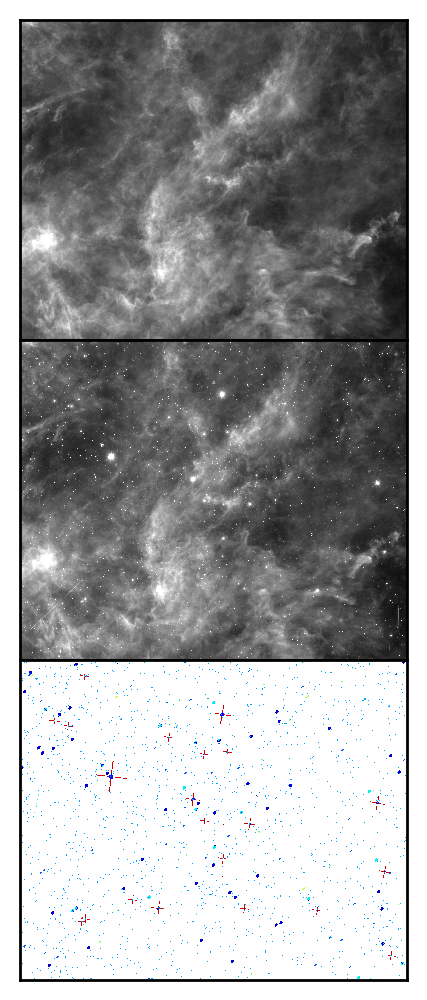}
\caption{\label{fig:tile} $4.3^{\circ} \times 3.8^{\circ}$ sub-region of 
          WSSA tile 51, at ($l$, $b$) $\approx$ ($266^{\circ}$, $-6^{\circ}$), 
          illustrating three of the extensions included in each of our tiles. 
          Top: primary extension, diffuse Galactic $W3$ emission. Middle: 
          `dirty' extension, including point sources, satellite streaks, and 
          SSOs. Bottom: colorscale rendering of `OR mask' extension. Red 
          crosses denote point source diffraction spikes.}
\end{center}
\end{figure}

\section{Final WSSA Tiles}
\label{sec:tiles}
In the following subsections we outline the final steps taken to convert our 
WSSA co-adds of $\S$\ref{sec:stack} into the data products we have released. 

\subsection{Zero Point}
\label{sec:zp}
Our mosaicking process only attempts to create a smooth image over each tile 
footprint by best matching each L1b exposure with its neighbors. However, this 
process does not ensure that neighboring tiles agree on their overlaps. 
Further, relics of the zodiacal light and Moon contamination that aren't 
perfectly removed by our SFD-based gradient correction can persist, leading to 
wandering of the tile zero points on relatively large scales of several 
degrees.

We investigated the possibility of obtaining a zero point by taking the
L1b zero point at face value, and then subtracting out models for the 
time-dependent foregrounds. Despite our best efforts to subtract the Moon-glow
model of $\S$\ref{sec:moon} and various customizations of the 
Kelsall zodiacal light model \citep{kelsall98}, such an approach ultimately 
yielded  a cosmetically unappealing result. Based on the Planck comparison 
we describe below, we estimate that, at high $|b|$, the true $W3$ Galactic 
emission is $\sim$0.5\% as bright as the $|\beta| = 90^{\circ}$ zodiacal light.
While the Kelsall model appeared reasonable at the several percent level, this 
remained inadequate due to the overwhelming brightness of the zodiacal light at
$12\mu$m.

We therefore resorted a zero point based on Planck \citep{planck13}. 
Henceforth, the term Planck will refer to Planck 857 GHz ($350\mu$m), 
corrected for zodiacal emission according to \cite{planck_zodi}. To ensure this
Planck map contained only diffuse Galactic emission, we also interpolated over 
the SFD compact source mask. The great advantage of Planck relative to other 
full-sky, mid-IR and far-IR datasets is the very faint zodiacal 
foreground. The basic strategy will be to replace modes on scales of several 
degrees in our WISE tiles with appropriately rescaled and smoothed Planck. 

In order to accomplish this goal, we must derive a map of the spatially
varying coefficient $f_{pw}$ that converts $350\mu$m emission to $12\mu$m 
emission. To derive $f_{pw}$ we correlate cleaned WISE with Planck on a 
bandpass between $4'$ and $15'$. We break each tile into many 
$0.5^{\circ} \times 0.5^{\circ}$ rectangular patches and perform a linear
least squares fit to the bandpass filtered WISE sub-image, where the model is 
bandpass Planck multiplied by $f_{pw}$,  plus a constant offset. For each such 
fit, we also compute the Pearson correlation coefficient $c$ of bandpass WISE 
versus bandpass Planck. We perform such fits over the full sky. For a variety 
of reasons, including poor signal to noise in Planck or WISE, or systematic 
issues in in Planck or WISE, $c$ is often closer to zero than to unity. In such
cases, the fit is driven to $f_{pw} \to 0$ for reasons that are instrumental 
rather than astrophysical. In order to obtain  a map of $f_{pw}$ on large 
scales, we make a cut on $c > 0.7$ and smooth the map of $f_{pw}$ at those 
locations to $15^{\circ}$. With WISE in $W3$ DN and Planck in MJy/sr $f_{pw}$, 
varies between 2.2 and 5.8. Converting WISE to MJy/sr according to 
\cite{cutri12} so that both data sets have the same units, this translates to
0.036 $ < f_{pw} < $ 0.095.

To replace the large scale modes of WISE with those of Planck, we 
apply a 12.5$'$ median filter to both data sets, and then smooth both to 
$2^{\circ}$. We inpainted over regions with strong molecular
emission, defined as Planck Type 3 CO $>$15 K~km/s \citep{planckco}. We 
subtract 2$^{\circ}$ WISE from each tile, and then add in smoothed Planck, 
multiplied by $f_{pw}$. In order to derive tiles free of edge effects 
from the smoothing process, we initially built our WSSA stacks on 
$14^{\circ} \times 14^{\circ}$ padded astrometric footprints.

Lastly, we convert the WISE tiles with Planck-based zero point
from $W3$ DN to MJy/sr according to the prescription of \cite{cutri12}, 
$\S$4.4h.

\subsection{Tile Format}
We release the 430 $W3$ WSSA tiles as multi-extension FITS files. Table 3 
lists the extensions included. Refer to the tile headers for further details.

\begin{deluxetable}{cl}
\tablewidth{0pc}
\tablecaption{WSSA Tile Extensions}
\tablehead{
\colhead{Extension} & \colhead{Description}
}
\startdata
0 & cleaned co-add \\
1 & dirty co-add \\
2 & integer coverage \\
3 & minimum value image \\
4 & maximum value image  \\
5 & AND bit-mask \\
6 & OR bit-mask \\
7 & transient artifact image 
\enddata
\label{table:ext}
\end{deluxetable}

\section{Discussion and Conclusion}
\begin{figure*} 
 \begin{center} 
  \includegraphics[width=7.0in]{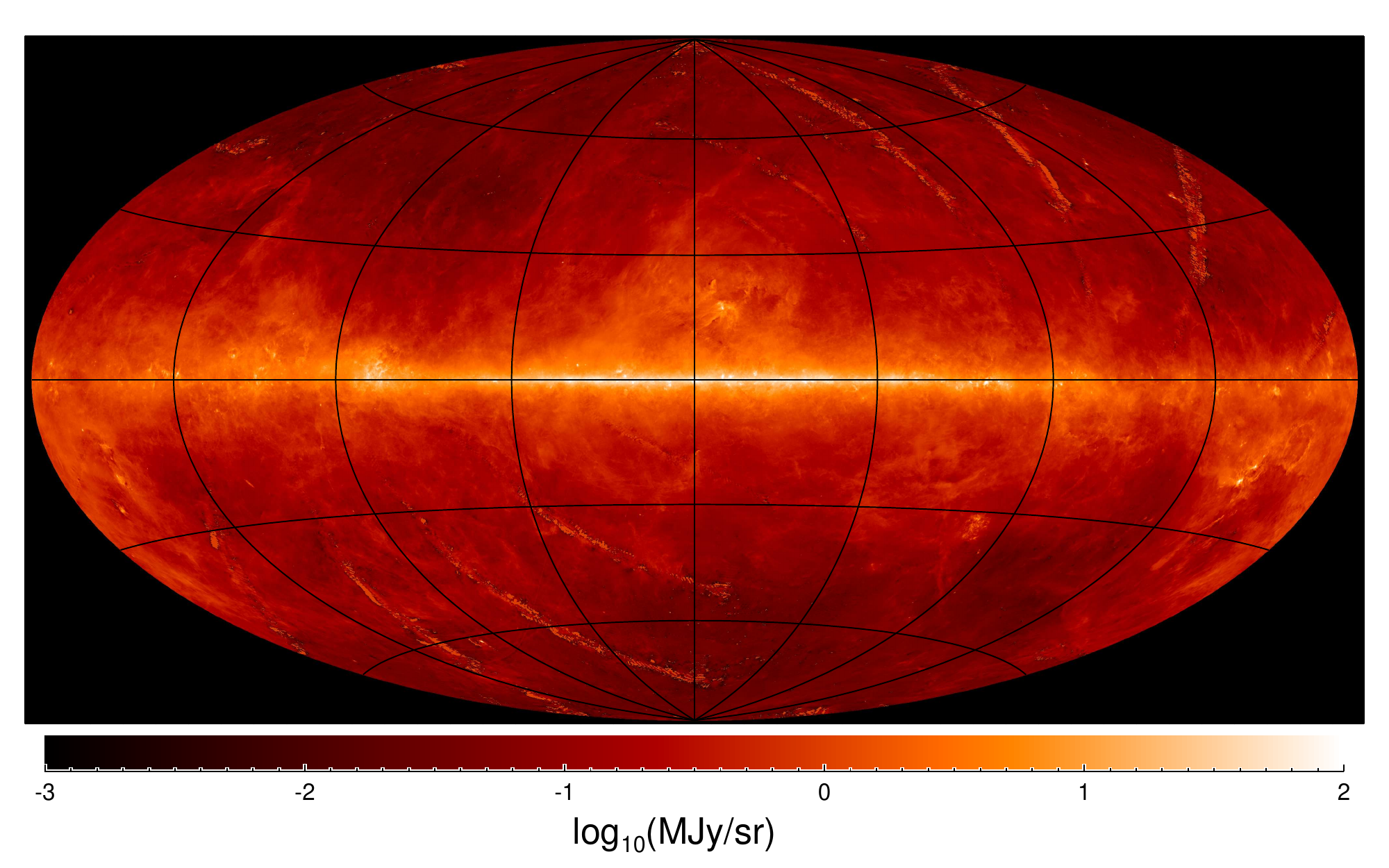}
  \caption{\label{fig:fullsky} Our full-sky, $12\mu$m map, binned to 7$'$ 
resolution, in Hammer-Aitoff projection, on a logarithmic stretch.}
 \end{center} 
\end{figure*}

\label{sec:disc}
We have described and released\footnote{See http://wise.skymaps.info.} a 
full-sky, $15''$ resolution map of diffuse
Galactic $12\mu$m emission based on WISE images. Indeed, WISE $W3$ 
provides valuable and striking small angular scale detail regarding the 
spatial structure of the ISM over much of the sky. We strongly encourage all 
interested astronomers to sample values from our mosaics, and also
visually inspect the co-adds. We anticipate this to be the first in a series 
of papers, throughout which we demonstrate that an enhanced-resolution dust map
based on WISE $W3$ better predicts extinction than does SFD, and ultimately 
release such a full-sky data product for accessing improved reddening 
estimates.

Figure \ref{fig:fullsky} shows the $12\mu$m map we have derived, binned 
to 7$'$ resolution. On large scales, the map looks very similar to the Planck
350$\mu$m data from which we derived our $12\mu$m zero-point. The notable 
exceptions are stripes of moon contamination, which dominate our map over 
$\sim$5\% of the sky.

We conclude by briefly pointing out aspects of the WISE All-sky release 
processing that turned out to be suboptimal for the purpose of mapping diffuse
emission, and finally compare our $12\mu$m map to those of DIRBE and \IRAS.

\subsection{All-sky Release Processing}
The ``dynamic calibration sky offset correction'' \citep[][$\S$4.4a]{cutri12}
introduced several problematic features into the L1b frames which we did not
address for this initial $12\mu$m tile release. In particular, 
long-term latents due to bright sources are mirrored in acausal fashion
to affect frames temporally \textit{preceding} bright source detection.
Examples of such features can be seen in Figure \ref{fig:stripe} as 
low-level, dark `divots' $\sim$1-2$'$ in size. These acausal latents are the 
dominant remaining imperfection in many regions of our co-adds with low
dust signal. In a related problem, the dynamic sky offset procedure can often 
assign up to $\sim$4\% of pixels in a given L1b image to exactly the median 
pixel value within the frame. Because these issues would be better addressed 
with lower-level data products than the L1b images, we have so far opted not 
to address them.

\subsection{DIRBE \& IRAS Comparison}
The relevant DIRBE data product to compare with our $W3$ map is the band 5
``Zodi-Subtracted Mission Average'' (ZSMA5). Because DIRBE pointed at a variety
of Solar elongations, including some $<$$90^{\circ}$, the DIRBE-observed
zodiacal light is even more pronounced and complex than that in $W3$. 
Arguably, the most prominent feature in ZSMA5 is 
the  \textit{residual} zodiacal light left over after subtracting the DSZA
model. The ecliptic plane is obvious in such residuals, and at high $|b|$
there remains a clear gradient of average flux with $\beta$, with magnitude
of $\sim$6$\times$$10^{-3}$ (MJy/sr)/deg, or $\sim$0.2 MJy/sr between
$|\beta|$$\sim$60$^{\circ}$ and $|\beta|$$\sim$90$^{\circ}$. By comparison, our
Planck-based far-IR zero point suggests that at $|b|$$\gtrsim$80$^{\circ}$, the
\textit{entire} non-IPD $12\mu$m signal is $\sim$0.03 MJy/sr. There is also a 
significant offset between ZSMA5 and our map in regions near 
the ecliptic poles, with amplitude $\sim$0.6 MJy/sr. While it might be 
interesting to investigate these differences, they are immaterial to our 
long-term goal of predicting dust extinction. Thanks to our far-IR zero point, 
our $W3$ map only shows traces of zodiacal light in rare circumstances, very 
near the ecliptic plane, where the zodi can vary strongly on scales smaller 
than $2^{\circ}$. Our $W3$ map is also superior to ZSMA5 in that point sources 
remain unsubtracted in the latter, impacting a substantial fraction of the 
sky. On the other hand, DIRBE band 5 is truly a full-sky map, whereas 
$\sim$5\% of diffuse $W3$ emission is badly contaminated by Moon-glow and/or 
Solar System planets.

\IRAS~12$\mu$m has a separate ``reject'' data set at low $|\beta|$ that is
not consistent with the remainder of their map. Thus, our $W3$ map
improves upon \IRAS~12$\mu$m by virtue of (1) its consistent, zodi-free zero 
point (2) its factor of 16 enhancement in angular resolution and (3) our 
removal and masking of compact sources.


\vspace{5mm}

We thank Roc Cutri for his guidance in making use of the WISE data 
products. We also thank Bruce Draine for his careful reading of an early 
version of this manuscript.

We acknowledge support of NASA grant NNX12AE08G for this research. This 
research was conducted with Government support under and awarded by DoD, 
Air Force Office of Scientific Research, National Defense Science and 
Engineering Graduate (NDSEG) Fellowship, 32 CFR 168a. This research made use of
the NASA Astrophysics Data System (ADS) and the IDL Astronomy User's Library 
at Goddard. \footnote{Available at \texttt{http://idlastro.gsfc.nasa.gov}}

This publication makes use of data products from the Wide-field Infrared 
Survey Explorer, which is a joint project of the University of California, Los 
Angeles, and the Jet Propulsion Laboratory/California Institute of Technology, 
funded by the National Aeronautics and Space Administration. This publication 
also makes use of data products from NEOWISE, which is a project of the Jet 
Propulsion Laboratory/California Institute of Technology, funded by the 
Planetary Science Division of the National Aeronautics and Space 
Administration. This publication makes use of data products from the Two Micron
All Sky Survey, which is a joint project of the University of Massachusetts and
the Infrared Processing and Analysis Center/California Institute of Technology,
funded by the National Aeronautics and Space Administration and the National
Science Foundation.

\bibliographystyle{apj}
\bibliography{wisetile.bib}

\end{document}